\documentclass
[floatfix,superscriptaddress,prl,twocolumn,noshowpacs,superbib]{revtex4}%
\usepackage{graphicx}
\usepackage{amsmath,amssymb,amsfonts}
\begin{document}
\title{Coulomb Correlations and the Wigner-Mott Transition}
\author{A.~Camjayi}
\affiliation{Department of Physics and Astronomy, Rutgers University, Piscataway, New
Jersey 08854, USA}
\affiliation{Departamento de F\'{\i}sica, Comisi\'on Nacional de Energ\'{\i}a At\'omica (CNEA), Avenida General Paz y Constituyentes, 1650 San Mart\'{\i}n, Argentina}
\author{K.~Haule}
\affiliation{Department of Physics and Astronomy, Rutgers University, Piscataway, New
Jersey 08854, USA}
\author{V.~Dobrosavljevi\'c}
\affiliation{Department of Physics and National High Magnetic Field Laboratory, Florida
State University, Tallahassee, FL 32310}
\author{G.~Kotliar}
\affiliation{Department of Physics and Astronomy, Rutgers University, Piscataway, New
Jersey 08854, USA}
\maketitle

\textbf{Strong correlation effects, such as a dramatic increase in the
  effective mass of the carriers of electricity, recently observed in
  the low density electron gas \cite{rpp1} have provided spectacular
  support for the existence of a sharp metal-insulator transitions in
  dilute two dimensional electron gases \cite{rmp251}.
%
%
  Here we show that strong correlations, normally expected only for
  narrow integer filled bands, can be effectively enhanced even far away
  from integer filling, due to incipient charge ordering driven by
  non-local Coulomb interactions. This general mechanism is
  illustrated by solving an extended Hubbard model using dynamical
  mean-field theory \cite{rmp13}. Our findings account for the key
  aspects of the experimental phase diagram, and reconcile the early
  view points of Wigner and Mott. The interplay of short range charge
  order and local correlations should result in a three peak structure
  in the spectral function of the electrons which should be observable
  in tunneling and optical spectroscopy.}

First indications of a two-dimensional metal-insulator transition
(2D-MIT) have emerged from transport studies, leading to a great deal
of controversy and debate \cite{rmp251}. Long-held beliefs
\cite{prl673} that even small amounts of impurities can destroy a
Fermi liquid at zero temperature were brought into question,
triggering renewed interest and activity. Careful theoretical work
\cite{prl016802} suggested that sufficiently strong interactions may
suppress weak localization tendencies and stabilize the metal at weak
disorder. More recent experiments
\cite{prl196404,prl126403,prl056805,rpp1,prl186404,prb241315} focused
on higher mobility (weaker disorder) samples, where advances in
experimental capabilities allowed precision measurements of the spin
susceptibility $\chi$ and the effective mass $m^{\ast}$. Within
experimental resolution, both quantities appear to diverge at the
critical density $n_{c}$, while the Wison ratio $\chi/m^{\ast} =
g^{\ast}$ appears to have a weaker density dependence. These findings,
which have been confirmed by several complementary experimental
methods \cite{prb205407,prl046409,prl036403}, are most clearly
pronounced in the cleanest samples, strongly suggesting that
interaction effects \cite{rpp1} - and not disorder - are the dominant
driving force for the 2D-MIT \cite{pankov2008}.

The divergence of the effective mass and spin susceptibility has been
observed in transition metal oxides near the density driven Mott
transition \cite{prl2126}, and in $^{3}$He monolayers near
solidification \cite{prl115301}.
%
%
For these materials, a description in
terms of an almost localized Fermi liquid, and the Brinkman-Rice
theory of the Hubbard model has been very successful \cite{rmp99}.
The similarity between the observation in oxides, $^3$He and 2D
electron gases (2DEG) suggests that we should think about the physics of the
2D-MIT as yet another example of the Hubbard-Mott phenomena
\cite{prb085317,jetpl377}.  Still, the situation relevant to the 2DEG
experiments corresponds to a nearly
empty conduction band - a regime very removed from near integer
filling where Mott-Hubbard physics has been successfully applied to
interpret experiments in $^3$He and transition metal oxides.

Another aspect of the Hubbard-Mott picture for 2D-MIT seems
equally troubling. Early theories of the Mott transition, using
the Gutzwiller variational approach \cite{rmp99}, did predict an
enhanced $m^{\ast}$ but a noncritical $ g^{\ast} $, as seen in
experiments. However more accurate calculations using
dynamical mean-field theory
(DMFT) \cite{rmp13} established that one generally should not
expect $\chi$ to diverge at the transition, 
but should instead saturate at a finite value $\chi
_{c}\sim1/J$, where $J$ is the (finite) superexchange interaction
characterizing the Mott insulating  phase of the lattice model in question. 
In this case one expects $ g^{\ast}
$ to gradually \textit{decrease and vanish} as the transition is
approached - in striking contrast to the 2DEG experiments.

Should one think of 2D-MIT as a manifestation of Mott physics - a
gradual conversion of the electrons into localized magnetic moments -
or does the explanation require a completely different physical
picture? In this paper we provide a simple answer to this important
question, and present detailed and careful model calculations to
support our view.
We envision that near the 2D-MIT the electron gas has short range
crystalline order, which we model with a tight binding Hamiltonian. 
The lattice sites represent the precursors, in the fluid phase, 
of vacancies and intersticials in the Wigner crystal phase.     
This is a lattice model at \textit{quarter filling} where the area of a cell containing
two lattice sites, corresponds to an area $\pi r_s^2 a_B^2$, containing one electron   
in the electron gas problem. Here $a_B$ is the Bohr radius and $r_s$ is the adimensional
ratio between Coulomb interaction and Fermi energy.
Since the system is not close to integer filling, the nonlocal
(inter-site) part of the Coulomb interaction can not be neglected, as
it induces significant charge correlations.
%
%
These in turn enhance the role of the short-range (on-site) part of
the Coulomb force, leading to strong correlation physics even far away
from integer filling. As the ratio of the Coulomb interactions
to the Fermi energy increases, the system
develops short range crystalline order which  in turn allows the  Hubbard interaction
to be effective resulting in  
the signatures of Mott localization.

\begin{figure}[!hb]
\begin{center}
\includegraphics[width=0.45\textwidth,clip]{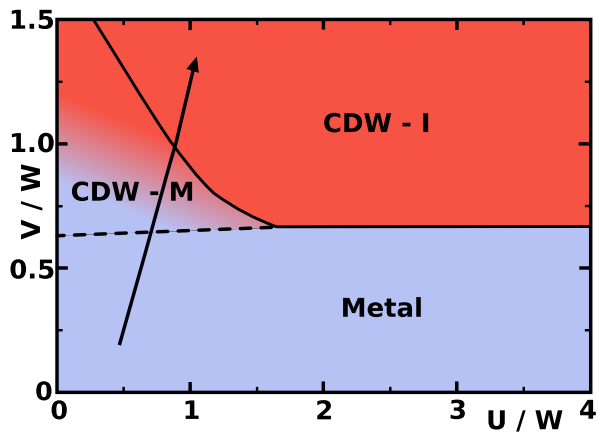}
\caption{\textbf{Phase diagram for the extended Hubbard model at quarter
filling.} The on-site Coulomb interaction $U$ and the inter-site
interaction $V$ are varied on $x$ and $y$-axis, respectively.
The temperature is held constant at $T=0.01$. $W$ is the half-bandwidth 
and the energy unit.
A typical trajectory relevant for the 2D-MIT is shown by the arrow
line. The following phases are found: CDW-M -- charge density wave
metallic phase, CDW-I -- charge density wave insulating phase.}
\label{fig:phdiag}%
\end{center}
\end{figure}

\textit{Model.} 
We neglect the effect of disorder and we focus on the extended Hubbard
model \cite{prl4046} as an effective Hamiltonian to describe the
physics of the 2D electron gas at low energies. This model contain in
addition to the usual on-site Hubbard $U$, a nearest neighbor (NN)
\textit{inter-site} repulsion $V$.
%
%
%
We envision that $V$ and $U$ are increasing functions of $r_s$ as
described in Fig.~\ref{fig:phdiag} (the arrow). We have
checked that all the qualitative features discussed in this paper do
not depend on the precise trajectory taken.
The spirit of our approach is similar to that of the almost localized
Fermi liquid framework \cite{rmp99,brinkman} which successfully
described key aspects of the physics of Helium near solidification.

The extended Hubbard model has been studied in detail using DMFT
\cite{rmp13} where the nonlocal part of the Coulomb interaction is
treated at the Hartree level. To incorporate the physics of Wigner
crystallization we consider a simple bipartite lattice, at quarter
filling, with a semi-circular density of states. The energy is
measured in units of the half-bandwidth $W=1$. To accurately solve the
DMFT equations at low temperatures ($T=0.01$) we utilize the
numerically exact continuous time quantum Monte Carlo algorithm as the
impurity solver \cite{prl076405}.


\begin{figure}[htb]
\begin{center}
\includegraphics[width=0.45\textwidth,clip]{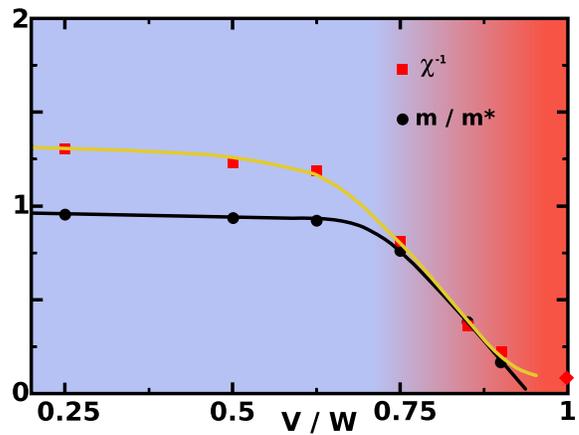}%
\caption{\textbf{Inverse effective mass and spin susceptibility.}
We plot these quantities along the arrow line in Fig.~\ref{fig:phdiag}.
Correlation effects are dramatically enhanced in presence
of charge ordering, while the system remains metallic. Black point
correspond to inverse effective mass $m/m^{\ast}$ and red squares
to inverse spin susceptibility $\chi^{-1}$. The lines are guides
to the eye.} \label{fig:chi-z}
\end{center}
\end{figure}

\textit{Wigner-Mott transition at quarter filling.} At quarter filling
and when the inter-site interaction $V$ vanishes, no insulating
solution is found even if the interaction parameter $U$ is arbitrarily
large. For $V>0$ charge ordering occurs. The DMFT phase diagram of
the system as a function of $U$ and $V$ is shown in
Fig.~\ref{fig:phdiag}.
%

The system goes from a weakly correlated Fermi liquid (small $U$ and
$V$), to a charge ordered Fermi liquid, to a Wigner-Mott insulator.
It is well known that broken symmetry phases in mean field theory,
are sometimes indicative of the onset of pronounced strong short
range order in the two dimensional system. Therefore we  cannot address with this approach
the possibility of the existence of  metallic charge ordered phase in the
electron gas \cite{prb075417}.


Remarkably strong correlation effects only emerge in the
intermediate regime, where charge ordering sets in. Increasing the charge
occupation on one of the two sublattices ($\langle
n_{A}\rangle\lessapprox1$), boosts the effects of the on-site
Coulomb repulsion $U$, and dramatically increases the correlation
effects. 
Hence, charge order leads to a 
dramatic increase of the effective mass and the spin
susceptibility, while the system remain metallic (Fig.~\ref{fig:chi-z}).
This behavior is strongly reminiscent of that found in the 2DEG experiments, 
where the mass enhancement is seen only in a narrow region 
preceding the  metal to insulator transition, but not at high  
densities, where $m/m^{\ast}\approx1$. 
%

The details of the magnetic interactions very close to the 2D-MIT, as well
as the different types of magnetic long range order in the insulator, depend
to some extent on the type of lattice used.
Note however that the enhancement of $\chi$ at the Wigner-Mott transition,
which is stronger at quarter filling than at half filling for the same
model, is a robust feature. 


   
\begin{figure}[htb]
\begin{center}
\includegraphics[width=0.45\textwidth,clip]{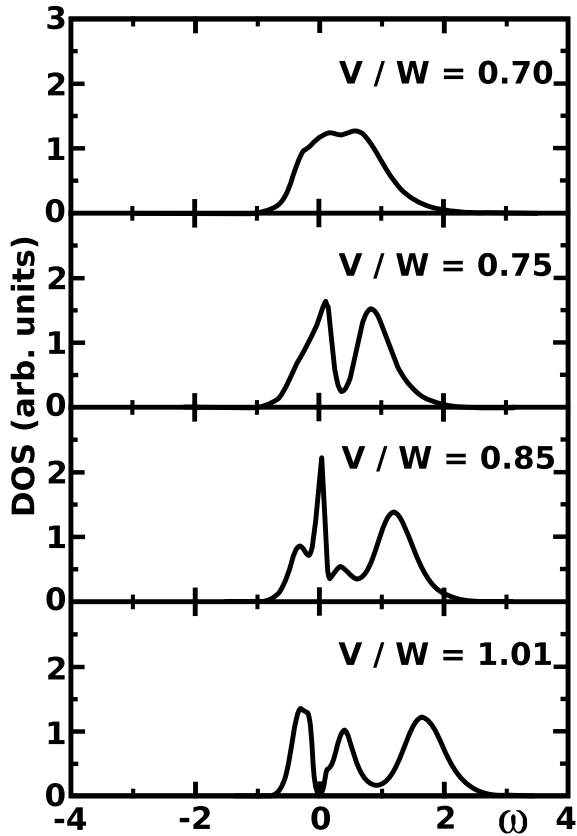}%
\caption{\textbf{Evolution of the density of states.} Correlations are more 
important as $U$ and $V$ increase along the path 
(see Fig.~\ref{fig:phdiag}), and a quasiparticle peak develops 
near the Fermi energy.}%
\label{fig:DOS}
\end{center}
\end{figure}

%
%
In the physical picture advocated in this approach the enhancement of
the effective mass is accompanied by the development of a
quasiparticle peak in the one particle density of states, as shown in
Fig.~\ref{fig:DOS}. The width of the quasiparticle peak is inversely
proportional to $m^*$.

\begin{figure}[htb]
\begin{center}
\includegraphics[width=0.45\textwidth,clip]{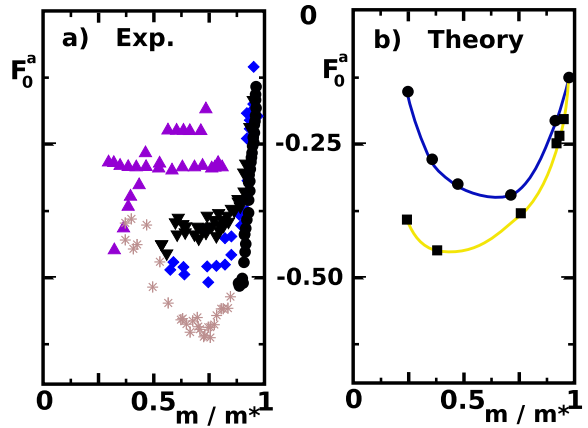}%
\caption{\textbf{Behavior of the Fermi liquid parameter $F_0^a$ versus
    inverse effective mass.} 
a) Experimental data showing the correlation between $F_0^a$ and
$m^*$. They are extracted from: \cite{prl196404} (black triangles),
 \cite{prl126403} (black circle), \cite{prl016405,prb045334} (blue
diamonds), \cite{prl046409} (violet triangles), and \cite{prl226401}
(maroon starts). 
b) Model calculations of the same quantity. The yellow curve
corresponds to the quarter filled case along the path marked by the
arrow in Fig.~\ref{fig:phdiag}. The blue curve shows the corresponding
Mott-Hubbard results at half-filling when the on-site $U$ is varying.
The Wigner-Mott model correctly captures the experimental trend. }
\label{fig:compar}
\end{center}
\end{figure}

A stringent test of our scenario is the relation between the
Fermi liquid parameter $F_0^a$ ($g^* = (1+F_0^a)^{-1}$) and the mass
enhancement. In Fig.~\ref{fig:compar} we show the behaviour of $F_0^a$
versus inverse mass $m/m^*$. Note the significant difference of this
quantity when measured in the vicinity of the Mott-Hubbard transition
or close to the Wigner-Mott transition at quarter filling. The latter
is strikingly similar to the available experimental data on the
2DEG.
The displayed experimental data was compiled from transport
measurements, magnetic measurements
\cite{prl016405,prb045334,prl226401} or both
\cite{prl126403,prl196404,prl046409} following a previous survey
\cite{nphys55} as a function of $r_s$.

\begin{figure}[!htb]
\begin{center}
\includegraphics[width=0.45\textwidth,clip]{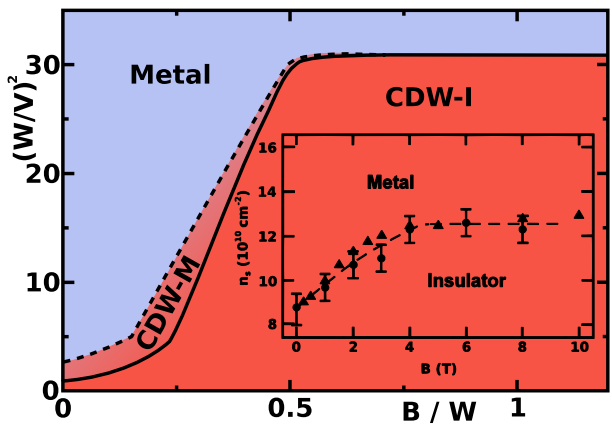}%
\caption{\textbf{Magnetic phase diagram.}
We evolve the phases along the path in Fig.~\ref{fig:phdiag} 
by applying a parallel magnetic field.
Field-driven localization is possible only sufficiently close to the
Wigner-Mott transition, within the correlated regime.  Inset:
experimental phase diagram. Adapted from References \cite{prl226403} and \cite{prl266402}.}%
\label{fig:mag}
\end{center}
\end{figure}

\textit{Magnetic field.} One of the most interesting features of
2D-MIT is the dramatic sensitivity of the correlated Fermi liquid
regime to the Zeeman (spin) splitting introduced by applying a
parallel magnetic field. Indeed, experiments demonstrated that the
heavy Fermi liquid can be effectively destroyed by applying a
parallel field, producing a spin-polarized insulating state above a 
``saturation field'' $B^{\ast}(n)$ of only a few Teslas. For a heavy 
Fermi liquid one expects $B^{\ast}\sim1/m^{\ast}$, and indeed 
experiments and our theory (see Fig.~\ref{fig:mag}) show that 
$B^{\ast}(n)$ $\sim(n-n_{c})$,
consistent with a singularly enhanced $m^{\ast}$ at the
transition. Such field-induced localization is only 
found in the correlated regime close enough to the transition.

This behavior is very hard to understand from the point of view of a
half-filled Hubbard model, since in this case sufficiently strong fields
always leads to insulating behavior. The field response we find at quarter
filling is dramatically different, as shown in Fig.~\ref{fig:mag}. A field-driven
localization transition is still found, but in contrast to the half-filled
case, it is restricted to the strongly correlated charge-ordered region; the
featureless Fermi liquid remains metallic even upon spin polarization. These
findings find surprising agreement with the experimentally established phase
diagram (Fig.~\ref{fig:mag}, inset).

The dependence of the effective mass on the applied magnetic field is
relatively weak, due to the presence of two competing effects. On one
hand the magnetic field locks the spin fluctuations, hence reducing
the entropy and the effective mass of the system.  On the other hand,
the magnetic field enhances the charge ordering, which in turn
produces a CDW coherence peak at the band edge, enhancing the density
of states.


\textit{Conclusions.} We presented the solution of a simple
microscopic model which correctly captures the qualitative behavior
near  a Mott  transition in the presence
of a charge ordered background  (Wigner-Mott transition). 
As in the
conventional Mott scenario, this transition features the formation of
a strongly correlated state on the metallic side.  The magnetic field
dependence proves to be dramatically different from what is expected
in a one band Hubbard model at half filling  but it is  in excellent
agreement with experiments on the 2DEG.

The current theory considers the on-site Coulomb repulsion at the single
site DMFT level 
and the nearest neighbor repulsion at the Hartree level. A better
treatment, which incorporates dynamical charge fluctuations, the
long range Coulomb interactions and short range correlation effects, 
is possible using extensions of DMFT.
%
We believe this treatment
will replace the long-ranged charge order, which we found on the
metallic side of the Wigner-Mott transition, by very strong short
range order characteristic of low density Coulomb systems.
While the key signatures of strong correlation will not be much
modified by these effects, the improved theory may allow a more
quantitative comparison with experiment.  Furthermore, disorder
effects in the strongly correlated regime need to be addressed. While
these directions remain interesting avenues for the future, we believe
that the essential new physics at the Wigner-Mott transition is
already captured within the present calculation.

The Wigner-Mott perspective, articulated in this letter, can be tested
by applying the new generation of experimental tools
used in the study of strongly correlated oxides to dilute electron gases in heterostructures.
Photoemission and scanning electron microscopy (STM) can probe 
the evolution of the one particle density of states shown in Fig.~\ref{fig:DOS}.
The characteristic spectral features associated
with the Mott phenomena, will also result in a reduction of the 
low frequency optical spectral weight, and this can be probed by microwave resonance methods.  
These experiments should be able to distinguish between a Wigner-Mott picture
of conversion of electronic degrees of freedom into local moments,
and the alternative perspective which views disorder as the
main driving force for the 2D-MIT \cite{science289}.

\end{document}